\documentclass[twocolumn,showpacs,aps]{revtex4}

\usepackage{graphicx}
\usepackage{latexsym}
\usepackage{amssymb}
\usepackage{color}
\usepackage[center]{subfigure}

\begin{document}

\newcommand{\bq}{\begin{equation}}
\newcommand{\eq}{\end{equation}}
\newcommand{\bqn}{\begin{eqnarray}}
\newcommand{\eqn}{\end{eqnarray}}
\newcommand{\nb}{\nonumber}
\newcommand{\lb}{\label}

\title{Vaidya Solution in General Covariant Ho\v{r}ava-Lifshitz Gravity 
with the Minimum Coupling and without Projectability: Infrared Limit}

\author{O. Goldoni $^{1,4}$ \footnote{{On leave from $^1$}}}
\email{otaviosama@gmail.com}
\author{M.F.A. da Silva $^{1}$}
\email{mfasnic@gmail.com}
\author{R. Chan $^{2}$}
\email{chan@on.br}
\author{G. Pinheiro $^{3}$}
\email{gpinheiro.fisica@gmail.com}
\affiliation{
$^{1}$ Departamento de F\'{\i}sica Te\'orica,
Instituto de F\'{\i}sica, Universidade do Estado do Rio de Janeiro,
Rua S\~ao Francisco Xavier 524, Maracan\~a
20550-900, Rio de Janeiro, RJ, Brasil.\\
$^{2}$ Coordena\c{c}\~ao de Astronomia e Astrof\'{\i}sica, 
Observat\'orio Nacional, Rua General Jos\'e Cristino, 77, S\~ao Crist\'ov\~ao  
20921-400, Rio de Janeiro, RJ, Brazil.
$^{3}$ 
Instituto Federal de Educa\c c\~ao, Ci\^encia e Tecnologia do Rio de Janeiro,
Rua Senador Furtado, 121/125, Maracan\~a, 20270-021, Rio de Janeiro\\
$^{4}$ GCAP-CASPER, Physics Department, Baylor University, Waco, TX 76798-7316, USA }

\date{\today}

\begin{abstract}
In this paper, we have studied nonstationary radiative spherically symmetric 
spacetime, in general covariant theory ($U(1)$ extension) of {the} Ho\v{r}ava-Lifshitz
gravity with the minimum coupling, in the post-newtonian approximation (PPN), 
without the projectability condition and in the infrared limit. 
The Newtonian prepotential $\varphi$ was assumed null.
We have shown that there is not the analogue of the Vaidya's solution in the 
Ho\v{r}ava-Lifshitz Theory (HLT) with the minimum coupling, as we know 
in the General Relativity Theory (GRT). 
\end{abstract}


\maketitle

\section{Introduction}

Since its publication, {the} Ho\v{r}ava-Lifshitz gravity theory (HLT) \cite{Horava}
\cite{Lifshitz} has draw a lot of interest
primarily due to its prospects to solve the perturbative non-renormalizability
of quantized standard Einstein gravity at the expense of breaking
relativistic invariance at very high energies, while restoring it at low
energies. After that, further deep
connections of {the} Ho\v{r}ava-Lifshitz formalism to {the} condensed matter physics were
discovered within the framework of {the} gauge/gravity duality (holography).

Unfortunately, the original Ho\v{r}ava formulation was contaminated by a number 
of serious problems which have caused the development of various non-trivial
modifications. One of the promising latter modifications is the one
without the so called projectability condition and with an additional Abelian
gauge symmetry, which in particular solves the problem with the unphysical
scalar graviton. Due to the extensive works in this area, we suggest to
the reader the references \cite{Visser}-\cite{Lin:2012bs}.

In order to establish the physical relevance of any modification of {the}
Ho\v{r}ava-Lifshitz gravity it is very important to check that the latter
reproduce the known physically feasible properties of {the} standard {Einstein's}
General Relativity (EGR) at low energies. Within this condition, we 
have checked whether the non-projectable version of {the} Ho\v{r}ava-Lifshitz gravity 
with the extra U(1) gauge symmetry contains in the low energy limit solutions of 
Vaidya-type \cite{Goldoni2014}. The answer was negative 
and, therefore, led us to the important conclusion that in order 
to establish consistency with {the} EGR at low energies the
gauge field associated with the extra U(1) gauge symmetry of the enlarged
non-projectable Ho\v{r}ava-Lifshitz gravity should have some interaction with 
the pure radiation matter generating the Vaidya spacetime geometry.

In a recent paper, Lin et al. (2014) \cite{Lin2014},
have proposed a universal coupling between the gravity
and matter in the framework of the Ho\v{r}ava-Lifshitz theory of gravity
with an extra U(1) symmetry for both the projectable and
non-projectable cases. Then, using this universal 
coupling they have 
studied 
the PPN approximations and they have obtained the 
PPN parameters in terms of the coupling constants of the theory. 

In this present work, using the results of Lin et al. (2014) \cite{Lin2014}
we have studied nonstationary radiative spherically 
symmetric spacetime, in {the} general covariant theory ($U(1)$ extension) of
{the} Ho\v{r}ava-Lifshitz gravity with a minimum coupling \cite{Lin2014}, 
in the
PPN approximation, without the projectability condition and in the infrared 
limit.  We will analyze if the Vaidya's spacetime can be described
as a null radiation fluid in the general
covariant HLT of gravity without the projectability condition \cite{ZWWS,ZSWW}.
In Section II we present a brief introduction to the HLT with the minimum
coupling \cite{Lin2014} considered here and {we} present the field equations of the HLT modified.
In Section III we show the Vaidya's spacetime, expressed in ADM decomposition\cite{ADM}.  
In Section IV we present the HLT equations in the PPN approximation, for the infrared limit. 
In Section V we analyze all the possible solution for 
the HLT field equations. In Section VI we discuss the results. Finally, in  
the Appendix A we present some quantities necessary in the HLT field equations 
with minimum coupling and without projectability \cite{Lin2014}.

\section{General Covariant Ho\v{r}ava-Lifshitz Gravity with Minimum Coupling and without Projectability}

In this section, we shall give a very brief introduction to the general 
covariant HLT gravity with the minimum coupling and without the projectability 
condition. For detail, we refer readers to \cite{ZWWS,ZSWW,Lin2014}.

The Arnowitt-Deser-Misner (ADM) form is given by \cite{ADM},
\bqn
ds^{2} &=& - N^{2}dt^{2} + g_{ij}\left(dx^{i} + N^{i}dt\right)
\left(dx^{j} + N^{j}dt\right), \nb\\
& & ~~~~~~~~~~~~~~~~~~~~~~~~~~~~~~  (i, \; j = 1, 2, 3),
\lb{ds2}
\eqn
where the nonprojectability condition imposes that $N=N(t,x^i)$.

In the work of Lin et al. (2014) \cite{Lin2014} it is proposed that, in the IR limit, it is possible to have matter fields universally
couple to the ADM components through the transformations
\bqn
\lb{eq8-1}
& & \tilde{N} = F N,\;\;\;
\tilde{N}^i = N^i + Ng^{ij} \nabla_j\varphi,\nb\\
&& 
\tilde{g}_{ij} = \Omega^2g_{ij},
\eqn
with
\bqn
\lb{eq8-1a}
& & F = 1 - a_1\sigma, \;\;\;
 \Omega = 1 - a_2\sigma,
\eqn
where
\bqn
\sigma &\equiv &\frac{A - {\cal{A}}}{N},\nb\\
{\cal{A}} &\equiv& - \dot{\varphi}  + N^i\nabla_i\varphi
+\frac{1}{2}N\left(\nabla^i\varphi\right)\left(\nabla_i\varphi\right),\nb\\
\eqn
and where $A$ and $\varphi$ 
are the the gauge field and the Newtonian prepotential,
respectively, 
and $a_1$ and $a_2$ are two arbitrary coupling constants. Note that by
setting the first terms in $F$ and $\Omega$ to unity, we have used the
freedom to rescale the units of time and space. We also have
\bq
\lb{eq8-2}
\tilde{N}_i =\Omega^2\left(N_i + N\nabla_i\varphi\right),\;\;\;
\tilde{g}^{ij} = \Omega^{-2}g^{ij}.
\eq

Considering the exposed before, the matter action can be written as
\bqn
\lb{eu7}
S_{m} &=& \int{dtd^3x \tilde{N}\sqrt{\tilde{g}}\;  \tilde{\cal{L}}_{m}\left(\tilde{N}, \tilde{N}_i, \tilde{g}_{ij}; \psi_{n}\right)},
\eqn
where $\psi_n$ collectively  stands for matter fields. One can then define the 
matter stress-energy in the ADM decomposition, with the minimum coupling. 
%

Thus, the total action of the theory can be written as,
\bqn 
\lb{TA}
S &=& \zeta^2\int dt d^{3}x  \sqrt{g}N \Big({\cal{L}}_{K} -
{\cal{L}}_{{V}} +  {\cal{L}}_{{A}}+ {\cal{L}}_{{\varphi}}  + {\cal{L}}_{S}+\nb\\
& &\frac{1}{\zeta^2} {\cal{L}}_{M}\Big), 
\eqn
where $g={\rm det}(g_{ij})$, $N$ is given in the equation (\ref{ds2}) and
\bqn \lb{2.5}
{\cal{L}}_{K} &=& K_{ij}K^{ij} -   \lambda K^{2},\nb\\
{\cal{L}}_{V} &=&  \gamma_{0}\zeta^{2}  -  \Big(\beta_0  a_{i}a^{i}- \gamma_1R\Big)
+ \frac{1}{\zeta^{2}} \Big(\gamma_{2}R^{2} +  \gamma_{3}  R_{ij}R^{ij}\Big)\nb\\
& & + \frac{1}{\zeta^{2}}\Bigg[\beta_{1} \left(a_{i}a^{i}\right)^{2} + \beta_{2} \left(a^{i}_{\;\;i}\right)^{2}
+ \beta_{3} \left(a_{i}a^{i}\right)a^{j}_{\;\;j} \nb\\
& & + \beta_{4} a^{ij}a_{ij} + \beta_{5}
\left(a_{i}a^{i}\right)R + \beta_{6} a_{i}a_{j}R^{ij} + \beta_{7} Ra^{i}_{\;\;i}\Bigg]\nb\\
& & +  \frac{1}{\zeta^{4}}\Bigg[\gamma_{5}C_{ij}C^{ij}  + \beta_{8} \left(\Delta{a^{i}}\right)^{2}\Bigg],\nb\\
{\cal{L}}_{A} &=&\frac{A}{N}\Big(2\Lambda_{g} - R\Big), \nb\\
{\cal{L}}_{\varphi} &=&  \varphi{\cal{G}}^{ij}\big(2K_{ij}+\nabla_i\nabla_j\varphi+a_i\nabla_j\varphi\big)\nb\\
& & +(1-\lambda)\Big[\big(\Delta\varphi+a_i\nabla^i\varphi\big)^2  
+2\big(\Delta\varphi+a_i\nabla^i\varphi\big)K\Big]\nb\\
& & +\frac{1}{3}\hat{\cal G}^{ijlk}\Big[4\left(\nabla_{i}\nabla_{j}\varphi\right) a_{(k}\nabla_{l)}\varphi \nb\\
&&  ~~ + 5 \left(a_{(i}\nabla_{j)}\varphi\right) a_{(k}\nabla_{l)}\varphi + 2 \left(\nabla_{(i}\varphi\right)a_{j)(k}\nabla_{l)}\varphi \nb\\
&& + 6K_{ij} a_{(l}\nabla_{k)}\varphi \Big], \nb \\
{\cal{L}}_S &=&\sigma a_S,
\eqn
where
\bqn
a_S=\sigma_1a_ia^i+\sigma_2a^i_i,
\eqn
\bq
\zeta^{2} = \frac{1}{16\pi G},
\eq
where $G$ denotes the Newtonian constant, 
${\cal{L}}_M$ is the Lagrangian of matter fields,
$\hat{\cal G}^{ijlk} =  g^{il}g^{jk} - g^{ij}g^{kl}$
\cite{Lin2013}. 
Here $\Delta \equiv g^{ij}\nabla_{i}\nabla_{j}$, $\Lambda_{g}$ is the 
cosmological 
constant, and all the coefficients, $ \beta_{n}$ and $\gamma_{n}$, are
dimensionless and arbitrary.  
{Note that if $a_1=a_2=\sigma_1=\sigma_2=0$ we recover the HLT without
any coupling with the matter \cite{Goldoni2014}}.

In order to be consistent 
with observations in the infrared limit
\cite{Lin2014}, we assume that
\bq
\beta_1=\beta_2=\beta_3=\beta_4=\beta_5=\beta_6=\beta_7=\beta_8=\beta_9=0,
\eq
\bq
\gamma_0=\gamma_2=\gamma_3=\gamma_4=\gamma_5=\gamma_6=\gamma_7=\gamma_8=\gamma_9=0,
\eq
implying that
\bq
\Lambda_g \equiv \frac{1}{2} \zeta^{2}\gamma_{0}=0.
\eq

For the PPN approximation in minimum coupling theory, we have
\bq
\beta_0=-2(\gamma_1+1),
\eq
\bq
a_1=a_2=0,
\eq
\bq
\sigma_1=0,
\eq
\bq
\sigma_2=4(1-a_1)=4.
\eq

Then, for the choice of the parameters above, we have
\bq
\tilde{N} = N,
\eq
\bq
\tilde{N}^i = N^i,
\eq

and

\bq
\tilde{g}_{ij} = g_{ij}.
\eq

$C_{ij}$ denotes the Cotton tensor, defined by
\bq
\lb{1.12}
C^{ij} = \frac{ {{e}}^{ikl}}{\sqrt{g}} \nabla_{k}\Big(R^{j}_{l} - \frac{1}{4}R\delta^{j}_{l}\Big),
\eq
with  $e^{123} = 1$.  Using the Bianchi identities, one can show that 
$C_{ij}C^{ij}$ can be written in terms of the five independent sixth-order 
derivative terms in the form
\bqn
\lb{1.13}
C_{ij}C^{ij}  &=& \frac{1}{2}R^{3} - \frac{5}{2}RR_{ij}R^{ij} + 3 R^{i}_{j}R^{j}_{k}R^{k}_{i}  +\frac{3}{8}R\Delta R\nb\\
& &  +
\left(\nabla_{i}R_{jk}\right) \left(\nabla^{i}R^{jk}\right) +   \nabla_{k} G^{k},
\eqn
where
\lb{1.14}
\bqn
G^{k}=\frac{1}{2} R^{jk} \nabla_j R - R_{ij} \nabla^j R^{ik}-\frac{3}{8}R\nabla^k R.
\eqn

The Ricci and Riemann tensors 
$R_{ij}$ and $R^{i}_{\;\; jkl}$  all refer to the 3-metric $g_{ij}$, with 
$R_{ij} = R^{k}_{\;\;ikj}$ and
\bqn \lb{2.6}
 R_{ijkl} &=& g_{ik}R_{jl}   +  g_{jl}R_{ik}  -  g_{jk}R_{il}  -  g_{il}R_{jk}\nb\\
 &&    - \frac{1}{2}\left(g_{ik}g_{jl} - g_{il}g_{jk}\right)R,\nb\\
K_{ij} &\equiv& \frac{1}{2N}\left(- \dot{g}_{ij} + \nabla_{i}N_{j} +
\nabla_{j}N_{i}\right),\nb\\
{\cal{G}}_{ij} &\equiv& R_{ij} - \frac{1}{2}g_{ij}R + \Lambda_{g} g_{ij},\nb\\
a_{i} &\equiv& \frac{N_{,i}}{N},\;\;\; a_{ij} \equiv \nabla_{j} a_{i},\nb\\
\eqn
where $N_i$ is defined in the ADM form of the
metric \cite{ADM}, given by equation (\ref{ds2}).

The variations of the action $S$ (\ref{TA}) with respect to $N$ and $N^{i}$ 
give rise to the Hamiltonian and momentum constraints,
\bqn \label{hami}
{\cal{L}}_K + {\cal{L}}_V + F_V-F_\varphi-F_\lambda+{\cal{H}}_S= 8 \pi G J^t,\;\;
\eqn
\bqn \label{mom}
&& M_S^i+\nabla_j \bigg\{\pi^{ij} -(1-\lambda)g^{ij}\big(\nabla^2\varphi+a_k\nabla^k\varphi\big)\nb\\
&& ~~~~~~~~~~~~~~~ - \varphi {\cal{G}}^{ij} - \hat{{\cal{G}}}^{ijkl} a_l \nabla_k \varphi\bigg\} = 8\pi G J^i, ~~~~
\lb{jui}
\eqn
where
\bqn
{\cal H}_S&=&\frac{2\sigma_1}{N}\nabla_i\left[a^i\left(A-{\cal A}\right)\right]-\frac{\sigma_2}{N}\nabla^2\left(A-{\cal A}\right)\nb\\          &&+\frac{1}{2}\nabla_j\varphi\nabla^j\varphi,\nb\\       
M_S^i&=&-\frac{1}{2}a_S\nabla^i\varphi, \nb\\                                    J^i &=& -N \frac{\delta {\cal{L}}_M}{\delta N_i},\;\;
J^t = 2 \frac{\delta (N {\cal{L}}_M)}{\delta N},\nb\\
\pi^{ij}&=&-K^{ij}+\lambda K g^{ij},
\eqn
with
and  $F_V$, $F_{\varphi}$ and $F_\lambda$ are given in the Appendix A.

Variations of $S$ with respect to $\varphi$ and $A$ yield, respectively,
\bqn \label{phi}
&& \frac{1}{2} {\cal{G}}^{ij} ( 2K_{ij} + \nabla_i\nabla_j\varphi  +a_{(i}\nabla_{j)}\varphi)\nb\\
&& + \frac{1}{2N} \bigg\{ {\cal{G}}^{ij} \nabla_j\nabla_i(N \varphi) - {\cal{G}}^{ij} \nabla_j ( N \varphi a_i)\bigg\}\nb\\
&& - \frac{1}{N} \hat{{\cal{G}}}^{ijkl} \bigg \{ \nabla_{(k} ( a_{l)} N K_{ij}) + \frac{2}{3} \nabla_{(k} (a_{l)} N \nabla_i \nabla_j \varphi)\nb\\
&& - \frac{2}{3} \nabla_{(j} \nabla_{i)} (N a_{(l} \nabla_{k)} \varphi) + \frac{5}{3} \nabla_j (N a_i a_k \nabla_l \varphi)\nb\\
&& + \frac{2}{3} \nabla_j (N a_{ik} \nabla_l \varphi)\bigg\}+\Sigma_S \nb\\
&& + \frac{1-\lambda}{N} \bigg\{\nabla^2  \left.[N (\nabla^2 \varphi + a_k \nabla^k \varphi)\right.] \nb\\
&& - \nabla^i [N(\nabla^2 \varphi + a_k \nabla^k \varphi) a_i] \nb\\
&&+ \nabla^2 (N K) - \nabla^i ( N K a_i)\bigg \}
 = 8 \pi G J_\varphi,
\eqn
where
\bqn \label{Sigma}
\Sigma_S & = & -\frac{1}{2N}\Bigg\{\frac{1}{\sqrt{g}}\frac{\partial}{\partial t}\left[\sqrt{g}a_S\right]\nb\\
&&-\nabla_k\left[\left(N^k+N\nabla^k\varphi\right)a_S\right]\Bigg\}
\eqn
and
\bqn \label{ja}
R-2 \Lambda_g-a_S= 8 \pi G J_A,
\eqn
where
\bqn
J_\varphi = -\frac{\delta {\cal{L}}_M}{\delta \varphi},\;\;\;
J_A= 2 \frac{\delta ( N {\cal{L}}_M)}{\delta A}.
\eqn

On the other hand, the variation of $S$ with respect to $g_{ij}$ yields the 
dynamical equations,
\bqn \label{dyn}
\frac{1}{\sqrt{g}N} \frac{\partial}{\partial t}\left(\sqrt{g} \pi^{ij}\right)+2(K^{ik}K^j_k-\lambda K K^{ij})\nb\\
-\frac{1}{2}g^{ij}{\cal{L}}_K+\frac{1}{N}\nabla_k (\pi^{ik}N^j+\pi^{kj}N^i-\pi^{ij}N^k)\nb\\
+F^{ij}-F^{ij}_S-\frac{1}{2}g^{ij}{\cal{L}}_S+F^{ij}_a-\frac{1}{2}g^{ij}{\cal{L}}_A+F^{ij}_\varphi\nb\\
-\frac{1}{N}(AR^{ij}+g^{ij}\nabla^2A-\nabla^j\nabla^iA)
=8\pi G \tau^{ij},\;\;\;\;\;\;
\eqn
where
\bqn
\lb{tauij}
\tau^{ij}&=&\frac{2}{\sqrt{g}N} \frac{\delta(\sqrt{g}N{\cal{L}}_M)}{\delta g_{ij}}, \nb\\
\eqn
and $F^{ij}$, $F^{ij}_S$, $F^{ij}_a$ and $F^{ij}_\varphi$ are given in the 
Appendix A.


\section{Vaidya's Spacetime}

The Vaidya's spacetime with an ingoing null dust usually written in the form 
\cite{BS09},
\bq
\lb{3.1}
ds^2 = - \left(1 - \frac{2m(v)}{r}\right) dv^2 + 2dvdr + r^2d\Omega^2,
\eq
where $d\Omega^2 \equiv d\theta^2 + \sin^2\theta d\phi^2$, and the 
corresponding energy-momentum tensor is given by
\bq
\lb{3.2}
T_{\mu\nu} = \rho(v,r)l_{\mu}l_{\nu},
\eq
with
\bq
\lb{3.3}
\rho =  \frac{2}{r^2}\frac{dm}{dv},\;\;\; l_{\mu} = - \delta_{\mu}^{v}.
\eq
Hereinafter, the Newtonian prepotential $\varphi$ is assumed null and $G=c=1$.

Introducing a time-like coordinate $t$ via the relation, $v = 2(t + r)$, the 
metric (\ref{3.1}) can be cast in the form,
\bqn
\lb{metric}
ds^2 &=& - \frac{r}{M}dt^2 + \frac{4M}{r}\left[dr + \left(1 - \frac{r}{2M}\right)dt\right]^2 \nb\\
&& + r^2d\Omega^2,
\eqn
where 
\bq
\lb{3.4}
M \equiv M(V) = 2m(v),\;\;\; V \equiv t + r. 
\eq
From equation (\ref{metric}), we immediately obtain 
\bqn
\lb{3.5a}
N &=& \sqrt{\frac{r}{M}},\;\;\; N^{i} = \left(1 - \frac{r}{2M}\right)\delta^{i}_{r},\nb\\
g_{rr} &=& \frac{4M}{r},\;\;\; g_{\theta \theta} = r^2, \;\;\;  g_{\phi \phi} = r^2\sin^2\theta,
\eqn
and
\bqn
\lb{3.5b}
N_{i} &\equiv& g_{ij}N^{j} =   -2 \left(1 - \frac{2M}{r}\right)\delta_{i}^{r},\nb\\
g^{rr} &=& \frac{r}{4M},\;\;\; g^{\theta\theta} = \frac{1}{r^2}, \;\;\;  g^{\phi\phi} = \frac{1}{r^2\sin^2\theta},\\
\rho &=& \frac{M^{*}}{2r^2},\;\;\; l_{\mu} = - 2\left(\delta^{t}_{\mu} + \delta^{r}_{\mu}\right),
\eqn
where ${M}^{*}  \equiv dM/dV$.  

Since $M=M(V)$, introducing another independent variable, $U=t-r$, we can 
find that 
\bq
M'=\dot M= \frac{1}{2}M^{*},
\lb{M*}
\eq
since $dM(V)/dU=0$.

Then, we find that the non null metric components are
\bqn
\lb{3.5c}
{}^{(4)}g_{tt} &=& -\left(N^2 - N_{i}N^{i}\right) = -\frac{4}{r}\left(M-r\right),\nb\\
{}^{(4)}g_{ti} &=& N_{i} = \frac{2}{r}\left(2M-r\right) \delta^r_i,\nb\\
{}^{(4)}g_{rr} &=& g_{rr}= \frac{4M}{r},\nb \\
{}^{(4)}g_{\theta\theta} &=& g_{\theta\theta}= r^2,\nb \\
{}^{(4)}g_{\phi\phi} &=& g_{\phi\phi}= r^2 \sin^2 \theta,\nb \\
{}^{(4)}g^{tt} &=& -\frac{1}{N^{2}} = -\frac{M}{r}  \nb\\
{}^{(4)}g^{ti} &=& \frac{N^{i}}{N^2} = \frac{M}{r}\left(1-\frac{r}{2M}\right) \delta^i_r,\nb\\
{}^{(4)}g^{rr} &=& 1-\frac{M}{r},\nb  \\
{}^{(4)}g^{\theta\theta} &=& \frac{1}{r^2},\nb  \\
{}^{(4)}g^{\phi\phi} &=& \frac{1}{r^2 \sin^2 \theta},\nb  \\
\eqn

\bqn
\lb{3.5d}
{}^{(4)}g^t_t &=& 1, \nb\\
{}^{(4)}g^r_r &=& 1, \nb \\
{}^{(4)}g^\theta_\theta &=& 1, \nb \\
{}^{(4)}g^\phi_\phi &=& 1. \nb \\
\eqn

For the projection tensor the non null components are
\bqn
\lb{3.5e}
{}^{(4)}h^r_t &=& 1-\frac{r}{2M}, \nb\\
{}^{(4)}h^r_r &=& 1, \nb \\
{}^{(4)}h^\theta_\theta &=& 1, \nb \\
{}^{(4)}h^\phi_\phi &=& 1. \nb \\
\eqn

Then, it can be shown that
\bqn
\lb{3.6}
n_{\mu} &=& N\delta^{t}_{\mu} = \sqrt{\frac{r}{M}},\nb\\
n^{\mu} &\equiv& {}^{(4)}g^{\mu\nu}n_{\nu} = -\frac{1}{N}\left(\delta^{\mu}_{t} - N^i\delta^{\mu}_{i}\right) \nb \\ 
&=& \sqrt{\frac{M}{r}}\left[-\delta^\mu_t+\left(1-\frac{r}{2M}\right) \delta^\mu_r\right], \\ 
h^\mu_\nu &=& ^{(4)}g^\mu_\nu + n^\mu n_{\nu}, \nb\\
h^r_t &=& 1-\frac{r}{2M}, \nb\\
h^r_r &=& 1, \nb\\
h^\theta_\theta &=& 1, \nb\\
h^\phi_\phi &=& 1, \\
J_i &=&  T_{\mu\nu}n^{\mu}h^{\nu}_{i}\nb \\
&=&\frac{1}{r^2}\sqrt{\frac{M}{r}}\frac{dM}{dV}\delta^r_i \nb \\
&=&\frac{1}{r^2}\sqrt{\frac{M}{r}}(\dot M + M')\delta^r_i,  \\
\lb{jdi}
\tau_{ij} &=&  T_{\mu\nu}h^{\mu}_{i}h^{\nu}_{j}\nb \\
&=& \frac{8M}{r^3}\frac{dM}{dV}\delta^r_i \delta^r_j\nb \\
&=&\frac{8M}{r^3}(\dot M + M')\delta^r_i\delta^r_j, \\ 
\eqn
where the prime and dot denotes the partial differentiation in relation to the
coordinate $r$ and $t$, respectively.

\section{Infrared Limit}

In the infrared limit we must have

\bq
J_t = -2 \rho.
\eq

Besides, hereinafter, we have assumed that $\lambda=1$.

Thus we have
\bqn
\lb{3.7}
K_{rr}&=&\frac{2 M' M r+M' r^2-2 \dot M M r-2 M^2-M r}{\sqrt{r/M} M r^2}, \nb\\
K_{\theta\theta}&=&\frac{r(2 M-r)}{2 \sqrt{r/M} M}, \nb\\
K_{\phi\phi}&=&\sin^2 \theta \frac{r (2 M-r)}{2 \sqrt{r/M} M}, \nb \\
K &=& \frac{2 M' M r+M' r^2-2 \dot M M r+6 M^2 -5 M r}{4 \sqrt{r/M} M^2 r}, \nb \\
R_{rr} &=& \frac{M' r-M}{M r^2}, \nb \\
R_{\theta\theta}&=&\frac{M' r^2+8 M^2-3 M r}{8 M^2}, \nb \\
R_{\phi\phi}&=&\sin^2 \theta \frac{M' r^2+8 M^2-3 M r}{8 M^2}, 
\eqn
\bq
\lb{Riccscalar}
R = \frac{M' r^2+4 M^2-2 M r}{2 M^2 r^2}, 
\eq
and
\bqn
{\cal{L}}_K &=& \frac{1}{16 M^3 r^3} \times[(\lambda-1)(-4 M'^2 M^2r^2 -4 M'^2M r^3-\nb\\
& &M'^2r^4+8 \dot M M' M^2r^2+4 \dot M M' M r^3+\nb\\
& &9M^2 r^2+8M' M^2 r^2- 4 \dot M M^2 r^2-\nb\\
& &4\dot M^2 M^2 r^2-36M^4+28M^3 r+2M' M r^3)-\nb\\
& &8(3\lambda+1)M^3 r(M' -\dot M)+\lambda(8M' M r^3-\nb\\
& &16\dot M M^2 r^2 +32M^3 r-16M^2 r^2)].
\eqn
\bqn
{\cal{L}}_V &=& \frac{1}{2 M^2 r^2} [M' r^2 \gamma_1+4 M^2-2 M r] 
\eqn
\bqn
F_V&=&\frac{1}{8 M^3 r}\left[4 M'' M r^2 \gamma_1+4 M'' M r^2-7 M'^2 r^2 \gamma_1-\right.\nb\\
& &\left. 7 M'^2 r^2+14 M' M r \gamma_1+14 M' M r-7 M^2 \gamma_1-\right.\nb\\
& &\left. 7 M^2\right],
\eqn
\bqn
{\cal H}_S &=& \frac{1}{8M^5 r^3}\times \left[\sqrt{\frac{r}{M}} A' M^4 r^2(4M' r-20 M)-\right.\nb\\
& &\left. 4 M''' M^2 r^4+30 M'' M' M r^4+8M'' M^3 r^2-\right.\nb\\
& &\left. 16 M'' M^2 r^3-35 M'^3 r^4-12M'^2 M^2 r^2+\right.\nb\\
& &\left. 40M'^2 M r^3-24 M' M^3 r+13 M' M^2 r^2-\right.\nb\\
& &\left.60M^4+6M^3 r\right].
\eqn

From equation (\ref{mom}) we have
\bqn
H&=&{\cal{L}}_K+{\cal{L}}_V+F_V+{\cal H}_S=8\pi J^t=\frac{1}{16 M^5 r^3}\times\nb \\
& &\left[4(\lambda-1)(-M'^2 M^4 r^2-M'^2 M^3 r^3+\right.\nb\\
& &\left. 2M'\dot M M^4 r^2+M'\dot M M^3 r^3+2M' M^4 r^2-\right. \nb\\
& &\left. \dot M^2 M^4 r^2-9M^6)+8(1+3\lambda)M^5 r(\dot M-M')+\right.\nb \\
& &\left. 8\sqrt{\frac{r}{M}}A' M^4 r^2(M' r-5M)-8M'' M^2 r^4+\right. \nb \\
& &\left. 60M'' M' M r^4 +8M'' M^3 r^4(\gamma_1+1)+\right.\nb\\
& &\left. 16M'' M^3 r^2-32M'' M^2 r^3-70M'^3 r^4-(14\gamma_1+\right. \nb \\
& &\left. \lambda)M'^2 M^2 r^4+(36\gamma_1+10\lambda)M' M^3 r^3+\right.\nb\\
& &\left. 4(1-5\lambda)\dot M M^4 r^2+2(30\lambda-14)M^5 r-\right. \nb \\
& &\left. 5(1+5\lambda)M^4 r^2-13M'^2 M^2 r^4-24M'^2 M^2 r^2+\right. \nb \\
& &\left. 8M'^2 M r^3+26M' M^3 r^3-48M' M^3 r+\right.\nb\\
& &\left. 26M' M^2 r^2+32M^5 r\gamma_1-30M^4 r^2\gamma_1-\right.\nb\\
& &\left. 120M^4+12M^3 r\right],
\eqn

\bqn
J_r &=&\frac{1}{8\sqrt{\frac{r}{M}}M^3 r^2}\times \nb\\
& &[(\lambda-1)(-4\dot{M}' M^2 r^2+4 M'' M^2 r^2+2 M'' M r^3-\nb\\
& &2 M'^2 M r^2-3 M'^2 r^3+2 M'\dot{M} M r^2+\nb\\
& &4 M' M^2 r+6 M' M r^2+5 M^2 r-18 M^3)+\nb\\
& &\dot{M} M^2 r(2\lambda+6)],
\lb{jdr}
\eqn

\bqn
J_A&=&\frac{1}{4 M^3 r^2}\left[2 M'' M r^3-3 M'^2 r^3+8 M' M r^2+ \right.\nb \\
& &\left. 8 M^3-7 M^2 r \right],
\lb{JA}
\eqn

\bqn
J_{\varphi}&=&\frac{1}{64 \sqrt{r/M} M^5 r^3}\times \nb \\
& &\left\{(\lambda-1)[16\dot{M}' M^2 r^3(M-M'r)+\right.\nb \\
& &\left. 4M^2 r^4(-2M''' M-M''' r+2\dot{M}'' M)+ \right.\nb \\
& &\left. 2M'' M R^4(10M' M+11M' r-2\dot{M} M)+\right.\nb \\
& &\left. M'^2 r^4(-10M' M-21M' r+10M\dot{M})-\right.\nb \\
& &\left. 4M'\dot{M} M^2 r^3]+\right.\nb\\
& &\left. \sqrt{\frac{r}{M}}[16M''' M^2 r^4(2M-r)-\right. \nb \\
& &\left. M'' M' M r^4(104r-176M)+\right.\nb \\
& &\left. M''M^2 r^3(80M-56r)+\right.\nb \\
& &\left. M'^3 r^4(168M-108r)+\right.\nb \\
& &\left. M'^2 M r^3(180r-216M)-\right.\nb \\
& &\left. 36M' M^2 r^2(2M+r)+\right.\nb \\
& &\left. 12M^3 r(10M-3r)]+\lambda(-28M'' M^3 r^3-\right.\nb \\
& &\left. 26M'' M^2r^4+22M'^2M^2r^2+51M'^2 M r^4-\right. \nb \\
& &\left. 30M' M^3 r^2-15M' M^2 r^3-6\dot{M} M^3 r^2+\right.\nb \\
& &\left. 18M^4 r-15 M^3 r^2)-16M'' M^3r^4+\right.\nb \\
& &\left. 12M'' M^3 r^3+34M'' M^2 r^4+24M'^2 M^2 r^4+\right. \nb \\
& &\left. 2M'^2 M^2 r^3-71M'^2 M r^4-32M' M^4 r-\right.\nb \\
& &\left. 48M' M^3 r^3-26M' M^3 r^2+59M' M^2 r^3+\right.\nb \\
& &\left. 32\dot{M} M^4 r-2\dot{M} M^3 r^2+32M^5+\right.\nb \\
& &\left. 24M^4 r^2+14M^4 r-9M^3 r^2\right\} .
\lb{Jphi}
\eqn

From the dynamical equation (\ref{dyn}) we have
\bq
D^{rr}=8 \pi \tau^{00}=\frac{d^{rr}}{128 \sqrt{r/M} M^{4} r^2}, 
\eq
where
\bqn
d^{rr}&=&\sqrt{\frac{r}{M}}\left[4(\lambda-1)\left(\dot{M}^2 M^2 r^2-4\dot{\dot{M}} M^3 r^2-\right.\right.\nb\\
& &\left.\left. 27 M^4+M'' M r^2(4 M^2-r^2)+4\dot{M}' M^2 r^3-\right.\right.\nb \\
& &\left.\left. 3 M'\dot{M} M r^3-3 M'^2 M r^2(M + r^2)+\right.\right.\nb \\
& &\left.\left. 6 M' M^2 r^2+2\dot{M} M' M^2 r^2\right)+116 M^3 r+\right.\nb \\
& &\left. M'^2 r^4(5\lambda+2\gamma_1-3)+8\dot{M} M^3 r(7\lambda+5)-\right.\nb\\
& &\left. 2 M' M r^3(2\gamma_1-\lambda+3)+\right.\nb\\
& &\left. 7 M^2 r^2(3-2\gamma_1-5\lambda)+4 M^3 r(8\gamma_1-21)-\right.\nb \\
& &\left. 4M^2 r(2M' M+\dot{M} r)(\lambda+3)\right]+\nb \\
& & 8A M^2 r(4M-r)-16A' M^2 r^3.
\eqn
\bq
D^{\theta\theta}=8 \pi \tau^{\theta\theta}=\frac{d^{\theta\theta}}{32 \sqrt{r/M} M^3 r^3}, 
\eq
where
\bqn
d^{\theta\theta}&=&\sqrt{\frac{r}{M}}\left[4(\lambda-1)\left({M'}^2 M r^2 (M+r)-\right.\right.\nb\\
& &\left.\left. \dot{M} M' M r^2 (2M+r)+10 M^3 r(\dot M - M')-\right.\right.\nb\\
& &\left.\left. 2M' M^2 r^2+\dot{M}^2 M^2 r^2+\dot{M} M^2 r^2+\right.\right.\nb\\
& &\left.\left. 9 M^4 + M^3 r\right)+\lambda(32 \dot{M}' M^3 r^2-\right.\nb\\
& &\left. 16\dot{M}^2 M^3 r^2-7 M'^2 r^4)+4 M'' M r^4 (\lambda+\gamma_1)-\right.\nb\\
& &\left. 16 M'' M^3 r^2- {M'}^2 r^4(10\gamma_1+3)+\right.\nb \\
& &\left. 2M' M r^3 (11\lambda+10\gamma_1-1)-\right.\nb \\
& &\left. 5 M^2 r^2 (3\lambda-2\gamma_1+1)\right]-8A'' M^2 r^4 +\nb\\
& & 4A' M r^3 (M' r-3M)+4 A M r^2(M' r-M),
\eqn
\bq
D^{\phi\phi}=8 \pi \tau^{\phi\phi}={D^{\theta\theta}}{\sin^2 \theta}. 
\eq

\section{Possible Solution}

We are looking for a HLT solution which is equivalent to the Vaidya's solution
in GRT. 
In the minimum coupling theory, we can see that $J_A=0$, since $\tilde N$ does
not depend on $A$ {(since $a_1=a_2=0$)} in the Vaidya's spacetime 
[see equation (4.11) in \cite{Lin2014}].
Then, since $J_A=0$, 
from equation (\ref{JA}) we have
\bq
2M r^3 M''-3r^3 M'^2 r^2+8M r^2 M'+8M^3 -7M^2 r=0.
\lb{JA1}
\eq

Using the equations $U=t-r$ and $V=t+r$ and the fact that $M=M(V)$, we can
write that
\bq
M'=\dot M=\frac{1}{2}M^*=\frac{1}{2}\frac{dM}{dV},
\eq
and
\bq
M''=\frac{1}{4} \frac{d^2M}{dV^2}.
\eq
Using these equations we can rewrite equation {the} (\ref{JA1}) as
\bqn
& &\frac{1}{16} M (V-U)^3 \frac{d^2M}{dV^2} - 
\frac{3}{32}(V-U)^3 \left(\frac{dM}{dV}\right)^2 + \nb \\
& &M(V-U)^2 \frac{dM}{dV} + 8M^3 - \frac{7}{2} M^2 (V-U)=0.\nb\\
\lb{JA2}
\eqn
Deriving {the} equation (\ref{JA2}) {three} times in relation to $U$ we get
\bq
-\frac{3}{8} M \frac{d^2M}{dV^2} +
\frac{9}{16} \left(\frac{dM}{dV}\right)^2  =0,
\lb{JA5}
\eq
whose solution is
\bq
M(V)=\frac{4}{(c_1 V + c_2)^2},
\lb{JA6}
\eq
where $c_1$ and $c_2$ are constants.
Substituting equation (\ref{JA6}) into (\ref{JA2}) we get
\bqn
& &6 c_1^2(V-U)^3-6c_1^2(c_1V+c_2)^2(V-U)^3- \nb \\
& &32c_1 (c_1V+c_2)^2(V-U)^2+512(c_1V+c_2)-\nb \\
& &56(c_1V+c_2)^2(V-U)=0.
\eqn
This equation can not be identically null 
since $c_1V+c_2 \ne 0$.

\section{Conclusion}

In this paper, we have analyzed nonstationary radiative spherically symmetric 
spacetime, in general covariant theory of Ho\v{r}ava-Lifshitz 
gravity with the minimum coupling, without the projectability condition, in the
PPN approximation and in the infrared limit. 
The Newtonian prepotential $\varphi$ was assumed null.

In our previously paper \cite{Goldoni2014} we {have} concluded that it 
does not exist a Vaidya's solution, as we know in {the} GRT.  
{The} unknowing coupling with the matter was the reason for {that} result.

Now we can conclude {at the present work} that, using the minimal coupling 
with matter, in the way proposed in \cite{Lin2014}, there is not a solution {in the HLT like} the solution of 
Vaidya, as we know in GRT. This result can suggest that this 
theory {needs} to be treated more carefully as a generalization of {the} GRT. 
{Therefore,} there is not any guarantee that starting from a given 
well-defined metric in {the}  GRT, this will lead to a solution in 
{the} HLT, whose infrared limit coincides {with} the corresponding 
solution of the GRT.

\begin{acknowledgements}
The financial assistance from FAPERJ/UERJ (MFAdaS) are gratefully acknowledged.
The author (RC) acknowledges the financial support from FAPERJ 
(no. E-26/171.754/2000, E-26/171.533/2002, E-26/170.951/2006, E-26/110.432\-/2009 
and E26/111.714/2010). The authors (RC and MFAdaS) also acknowledge the 
financial support from Conselho Nacional de Desenvolvimento Cient\'ifico e
Tecnol\'ogico - CNPq - Brazil (no. 450572/2009-9, 301973/2009-1 and 
477268\-/2010-2). The author (MFAdaS) also acknowledges the financial support 
from Financiadora de Estudos e Projetos - FINEP - Brazil (Ref. 2399/03).
We also would like to thank Dr. Anzhong Wang for helpful discussions and
comments about this work.
\end{acknowledgements}

\section{Appendix A: Definition of $F^{ij}$, $F_S^{ij}$, $F^{ij}_a$, $F^{ij}_\varphi$ and $F_S^{ij}$}

The quantities $F^{ij}$, $F_S^{ij}$, $F^{ij}_a$, $F^{ij}_\varphi$ and 
$F_S^{ij}$ are given by

\bqn
F^{ij}&=&\frac{1}{\sqrt{g}N}\frac{\delta (-\sqrt{g}N {\cal{L}}_V^R)}{\delta g_{ij}}\nb\\
&=& \sum_{s=0}\hat{\gamma}_s\zeta^{n_s}(F_s)^{ij},\\
F_S^{ij}&=&-\sigma \left(\sigma_1a^ia^j+\sigma_2a^{ij}\right)\nb\\
&&+\frac{a_S}{2}\left[(\nabla^i\varphi)(\nabla^j\varphi)+2\frac{N^{(i}\nabla^{j)}\varphi}{N}\right] \nb\\                              &&+\frac{\sigma_2}{N}\nabla^{(i}[a^{j)}(A-{\cal
                    A})]-g^{ij}\frac{\sigma_2}{2N}\nabla^{k}[a_k(A-{\cal
                    A})],\nb\\
F^{ij}_a&=&\frac{1}{\sqrt{g}N}\frac{\delta (-\sqrt{g}N {\cal{L}}_V^a)}{\delta g_{ij}}\nb\\
& =& \sum_{s=0}\beta_s\zeta^{m_s}(F_s^a)^{ij},\\
F^{ij}_\varphi&=&\frac{1}{\sqrt{g}N}\frac{\delta (-\sqrt{g}N {\cal{L}}_V^\varphi)}{\delta g_{ij}}\nb\\
&=& \sum_{s=0}\mu_s(F_s^\varphi)^{ij},\nb\\
F_S^{ij}&=&-\sigma \left(\sigma_1a^ia^j+\sigma_2a^{ij}\right)\nb\\
                &&+\frac{a_S}{2}\left[(\nabla^i\varphi)(\nabla^j\varphi)+2\frac{N^{(i}\nabla^{j)}\varphi}{N}\right] \nb\\
                    &&+\frac{\sigma_2}{N}\nabla^{(i}[a^{j)}(A-{\cal
                    A})]-g^{ij}\frac{\sigma_2}{2N}\nabla^{k}[a_k(A-{\cal
                    A})],\nb\\
\eqn
with
\bqn
\hat{\gamma}_s &=& \left(\gamma_0, \gamma_1, \gamma_2, \gamma_3, \frac{1}{2}\gamma_5, -\frac{5}{2}\gamma_5, 3\gamma_5, \frac{3}{8}\gamma_5, \gamma_5, \frac{1}{2}\gamma_5\right), \nb\\
n_s &=& (2, 0, -2, -2, -4, -4, -4, -4, -4, -4),\nb\\
m_s&=& (0, -2,-2,-2, -2, -2, -2, -2, -4 ), \nb\\
\mu_s &=& \left(2, 1, 1, 2, \frac{4}{3}, \frac{5}{3}, \frac{2}{3}, 1-\lambda, 2-2 \lambda\right).
\eqn

Thus, $F_V,\;F_\varphi$ and $F_\lambda$ are given, respectively, by
\bqn\label{a1}
F_V &=&  \beta_0 ( 2 a_i^i + a_i a^i) - \frac{\beta_1}{\zeta^2} \Bigg[3 (a_i a^i)^2 + 4 \nabla_i (a_k a^k a^i)\Bigg]\nb\\
    &&  +\frac{\beta_2}{\zeta^2}\Bigg[ (a_i^i)^2 + \frac{2}{N} \nabla^2 (N a_k^k)\Bigg]\nb\\
    && + \frac{\beta_3}{\zeta^2}\Bigg[ - (a_i a^i) a_j^j - 2 \nabla_i (a_j^j a^i) + \frac{1}{N} \nabla^2 (N a_i a^i)\Bigg]\nb\\
    &&+ \frac{\beta_4}{\zeta^2}\Bigg[ a_{ij} a^{ij} + \frac{2}{N} \nabla_j \nabla_i (N a^{ij})\Bigg]\nb\\
      && + \frac{\beta_5}{\zeta^2}\Bigg[- R (a_i a^i) - 2 \nabla_i (R a^i)\Bigg]\nb\\
      &&+ \frac{\beta_6}{\zeta^2}\Bigg[- a_i a_j R^{ij} - \nabla_i (a_j R^{ij})-\nabla_j (a_i R^{ij})\Bigg]\nb\\
      && +  \frac{\beta_7}{\zeta^2}\Bigg[ R a^i_i + \frac{1}{N} \nabla^2 (NR)\Bigg]\nb\\
      &&+ \frac{\beta_8}{\zeta^4}\Bigg[(\Delta a^i)^2 - \frac{2}{N} \nabla^i [\Delta (N \Delta a_i)]\Bigg],
\eqn
\bqn\label{a2}
F_\varphi &=& -  {\cal{G}}^{ij}\nabla_i \varphi \nabla_j \varphi, - \frac{2}{N} \hat{{\cal{G}}}^{ijkl} \nabla_l (N K_{ij} \nabla_k \varphi),\nb\\
        &&  - \frac{4}{3}\Bigg[ \hat{{\cal{G}}}^{ijkl} \nabla_l (\nabla_k \varphi \nabla_i \nabla_j \varphi)\Bigg]\nb\\
         &&+ \frac{5}{3}\Bigg[ -  \hat{{\cal{G}}}^{ijkl} [(a_i \nabla_j \varphi) (a_k \nabla_l \varphi)+\nabla_i(a_k\nabla_j\varphi\nabla_l\varphi)\nb\\
         && +\nabla_k(a_i\nabla_j\varphi\nabla_l\varphi)]\Bigg]\nb\\
         &&+ \frac{2}{3}\Bigg[ \hat{{\cal{G}}}^{ijkl}[a_{ik} \nabla_j \varphi \nabla_l \varphi + \frac{1}{N} \nabla_i\nabla_k (N\nabla_j\varphi\nabla_l\varphi)]\Bigg],\nb\\
\eqn
\bqn\lb{a3a}
F_\lambda &=& (1-\lambda) \Bigg\{(\nabla^2 \varphi + a_i \nabla^i \varphi)^2 - \frac{2}{N} \nabla_i (NK\nabla^i \varphi)\nb\\
           && - \frac{2}{N} \nabla_i [N (\nabla^2 \varphi + a_i \nabla^i \varphi) \nabla^i \varphi]\Bigg\}\label{a3}.
\eqn

$\left(F_n\right)_{ij}$, $\left(F^{a}_{s}\right)_{ij}$ and $\left(F^{\varphi}_{q}\right)_{ij}$, defined in equation (\ref{tauij}),  are given, respectively, by
\bqn
(F_0)_{ij} &=& -\frac{1}{2}g_{ij},\nb\\
(F_1)_{ij} &=& R_{ij}-\frac{1}{2}Rg_{ij}+\frac{1}{N}(g_{ij}\nabla^2 N-\nabla_j\nabla_i N),\nb\\
(F_2)_{ij} &=& -\frac{1}{2}g_{ij}R^2+2RR_{ij}\nb\\
             &&  +\frac{2}{N}\left[g_{ij}\nabla^2(NR)-\nabla_j\nabla_i(NR)\right],\nb\\
(F_3)_{ij} &=& -\frac{1}{2}g_{ij}R_{mn}R^{mn}+2R_{ik}R^k_{j}\nb\\
             &&  +\frac{1}{N}\Big[- 2\nabla_k\nabla_{(i}(NR_{j)}^k)\nb\\
             &&  +\nabla^2(NR_{ij})+g_{ij}\nabla_m\nabla_n(NR^{mn})\Big], \nb\\
(F_4)_{ij} &=&  -\frac{1}{2}g_{ij}R^3+3R^2R_{ij}\nb\\
             &&   +\frac{3}{N}\Big(g_{ij}\nabla^2-\nabla_j\nabla_i\Big)(NR^2),\nb\\
(F_5)_{ij} &=& -\frac{1}{2}g_{ij}RR_{mn}R^{mn}\nb\\
             &&+R_{ij}R_{mn}R^{mn}+2RR_{ik}R^k_{j}\nb\\
             &&  +\frac{1}{N}\Big[g_{ij}\nabla^2(NR_{mn}R^{mn})\nb\\
             &&-\nabla_j\nabla_i(NR_{mn}R^{mn})\nb\\
             &&  +\nabla^2(NRR_{ij})+g_{ij}\nabla_m\nabla_n(NRR^{mn})\nb\\
             &&  -2\nabla_m\nabla_{(i}(R^m_{j)}NR)\Big], \nb\\
(F_6)_{ij} &=& -\frac{1}{2}g_{ij}R^m_nR^n_lR^l_m+3R^{mn}R_{mi}R_{nj}\nb\\
             &&  +\frac{3}{2N}\Big[g_{ij}\nabla_m\nabla_n(NR^m_aR^{na}) \nb\\
             &&+ \nabla^2(NR_{mi}R^m_j)
              -2\nabla_m\nabla_{(i}(NR_{j)n}R^{mn})\Big], \nb\\
(F_7)_{ij} &=& -\frac{1}{2}g_{ij}R\nabla^2R+R_{ij}\nabla^2R+R\nabla_i\nabla_j R\nb\\
             &&  +\frac{1}{N}\Big[g_{ij}\nabla^2(N\nabla^2R)-\nabla_j\nabla_i(N\nabla^2R)\nb\\
             &&+R_{ij}\nabla^2(NR)
              +g_{ij}\nabla^4(NR)-\nabla_j\nabla_i (\nabla^2 (NR))\nb\\
             &&  - \nabla_{(j}(NR\nabla_{i)}R)+\frac{1}{2}g_{ij}\nabla_k(NR\nabla^kR)\Big], \nb\\
(F_8)_{ij} &=& -\frac{1}{2}g_{ij}(\nabla_mR_{nl})^2 + 2 \nabla^mR^n_i\nabla_mR_{nj}\nb\\
             &&  +\nabla_iR^{mn}\nabla_jR_{mn}+\frac{1}{N}\Big[2 \nabla_n\nabla_{(i}\nabla_m(N\nabla^mR^n_{j)})\nb\\
             &&  -\nabla^2\nabla_m(N\nabla^mR_{ij})-g_{ij}\nabla_n\nabla_p\nabla_m(N\nabla^mR^{np})\nb\\
             &&  -2\nabla_m(NR_{l(i}\nabla^mR^l_{j)})-2\nabla_n(NR_{l(i}\nabla_{j)}R^{nl})\nb\\
             &&  +2\nabla_k(NR^k_l\nabla_{(i}R^l_{j)})\Big], \nb\\
(F_9)_{ij} &=& -\frac{1}{2} g_{ij} a_k G^k+\frac{1}{2} \Big[a^k R_{k(j} \nabla_{i)} R + a_{(i} R_{j)k} \nabla^k R\Big]\nb\\
           &&-a_kR_{mi}\nabla_jR^{mk}-a^kR_{n(i}\nabla^nR_{j)k}\nb \\
           &&-\frac{1}{2}\Big[a_iR^{km}\nabla_mR_{kj}+a_jR^{km}\nabla_mR_{ki}\Big]\nb\\
           &&-\frac{3}{8}a_{(i}R\nabla_{j)}R+\frac{3}{8}\Bigg\{R\nabla_k(Na^k)R_{ij}\nb\\
           &&+g_{ij}\nabla^2\Big[R\nabla_k(Na^k)\Big]-\nabla_i\nabla_j\Big[R\nabla_k(Na^k)\Big]\Bigg\}\nb\\
           &&+\frac{1}{4N} \Bigg\{- \frac{1}{2}\nabla^m \Big[\nabla_{(i} Na_{j)}\nabla_m R+\nabla_{(i}(\nabla_{j)}R) Na_m\Big]\nb\\
           &&+\nabla^2 (N a_{(i}\nabla_{j)}R)+g_{ij} \nabla^m\nabla^n (Na_m\nabla_nR)\nb\\
           && +\nabla^m\Big[\nabla_{(i} (\nabla_{j)} R^k_m) Na_k+\nabla_{(i}(\nabla_m R^k_{j)})Na_k\Big]\nb\\
           &&-2\nabla^2(Na_k\nabla_{(i} R^k_{j)})-2g_{ij}\nabla^m\nabla^n(Na_k\nabla_{(n}R_{m)}^k)\nb\\
           &&- \nabla^m \Big[\nabla_i\nabla_p(Na_jR_m^p+Na_mR_j^p)\nb\\
           &&+\nabla_j\nabla_p(Na_iR_m^p+Na_mR_i^p)\Big]\nb\\
           &&+2\nabla^2\nabla_p(Na_{(i}R_{j)}^p)\nb\\
           && +2g_{ij}\nabla^m\nabla^n \nabla^p(Na_{(n}R_{m)p})\Bigg\}, \nb\\
\eqn

\bqn
(F_0^a)_{ij} &=&  -\frac{1}{2} g_{ij} a^k a_k +a_i a_j, \nb\\
(F_1^a)_{ij} &=&  -\frac{1}{2} g_{ij} (a_k a^k)^2 + 2 (a_k a^k) a_i a_j,\nb\\
(F_2^a)_{ij} &=&  -\frac{1}{2} g_{ij} (a_k^k)^2 + 2 a_k^k a_{ij}\nb\\
             &&   - \frac{1}{N} \Big[2 \nabla_{(i} (N a_{j)} a_k^k) - g_{ij} \nabla_\alpha (a_\alpha N a_k^k)\Big],\nb\\
(F_3^a)_{ij} &=&   -\frac{1}{2} g_{ij} (a_k a^k) a_\beta^\beta + a^k_k a_ia_j + a_k a^k a_{ij}\nb\\
             &&   - \frac{1}{N} \Big[ \nabla_{(i} (N a_{j)} a_k a^k) - \frac{1}{2} g_{ij} \nabla_\alpha (a_\alpha N a_ka^k)\Big],\nb\\
(F_4^a)_{ij} &=&  - \frac{1}{2} g_{ij} a^{mn} a_{mn} + 2a^k_i a_{kj} \nb\\
             &&   - \frac{1}{N} \Big[\nabla^k (2 N a_{(i} a_{j)k} - N a_{ij} a_k)\Big], \nb\\
(F_5^a)_{ij} &=&  -\frac{1}{2} g_{ij} (a_k a^k ) R + a_i a_j R + a^k a_k R_{ij} \nb\\
              &&  + \frac{1}{N} \Big[ g_{ij} \nabla^2 (N a_k a^k) - \nabla_i \nabla_j (N a_k a^k)\Big], \nb\\ 
(F_6^a)_{ij} &=&   -\frac{1}{2} g_{ij} a_m a_n R^{mn} +2 a^m R_{m (i} a_{j)} \nb\\
              &&  - \frac{1}{2N} \Big[ 2 \nabla^k \nabla_{(i} (a_{j)} N a_k) - \nabla^2 (N a_i a_j) \nb\\
               && - g_{ij} \nabla^m \nabla^n (N a_m a_n)\Big], \nb\\ 
(F_7^a)_{ij} &=&  -\frac{1}{2} g_{ij} R a_k^k + a_k^kR_{ij} + R a_{ij} \nb\\
             &&   + \frac{1}{N} \Big[ g_{ij} \nabla^2 (N a_k^k) - \nabla_i \nabla_j (N a_k^k) \nb\\
             &&  - \nabla_{(i} (N R a_{j)}) + \frac{1}{2} g_{ij} \nabla^k (N R a_k)\Big], \nb\\
(F_8^a)_{ij} &=&  -\frac{1}{2} g_{ij} (\Delta a_k)^2 + (\Delta a_i) (\Delta a_j) + 2 \Delta a^k \nabla_{(i} \nabla_{j)} a_k \nb\\
             &&   + \frac{1}{N} \Big[\nabla_k [a_{(i} \nabla^k (N \Delta a_{j)}) + a_{(i} \nabla_{j)} (N \Delta a^k)\nb\\
             &&   - a^k \nabla_{(i} (N \Delta a_{j)}) + g_{ij} N a^{\beta k} \Delta a_\beta  - N a_{ij} \Delta a^k ]\nb\\
             &&  -  2 \nabla_{(i} (N a_{j)k} \Delta a^k)\Big],
\eqn

\bqn
(F_1^\varphi)_{ij} &=&     -\frac{1}{2} g_{ij} \varphi {\cal{G}}^{mn}K_{mn}\nb\\
                   &&  + \frac{1}{2\sqrt{g} N}  \partial_t (\sqrt{g} \varphi {\cal{G}}_{ij}) -
                                                                                2 \varphi K_{(i}^\nu R_{j) \nu} \nb\\
                   &&       + \frac{1}{2} \varphi (K R_{ij} + K_{ij} R - 2 K_{ij} \Lambda_g ) \nb\\
                   &&      + \frac{1}{2N} \bigg\{{\cal{G}}_{ij} \nabla^k (\varphi N_k) - 2 {\cal{G}}_{k(i} \nabla^k (N_{j)} \varphi)\nb\\
                   &&      +  g_{ij}\nabla^2 (N \varphi K) - \nabla_i \nabla_j (N \varphi K) \nb\\
                   &&+ 2  \nabla^k \nabla_{(i} (K_{j)k} \varphi N),\nb\\
                   &&      - \nabla^2 (N \varphi K_{ij}) - g_{ij} \nabla^\alpha \nabla^\beta (N \varphi K_{\alpha \beta})\bigg\}, \nb\\
(F_2^\varphi)_{ij} &=&   - \frac{1}{2}g_{ij} \varphi {\cal{G}}^{mn} \nabla_m\nabla_n \varphi \nb\\
                   &&     - 2 \varphi \nabla_{(i} \nabla^k R_{j)k} + \frac{1}{2} \varphi (R- 2 \Lambda_g) \nabla_i \nabla_j \varphi \nb\\
                   &&     - \frac{1}{N} \bigg\{- \frac{1}{2} (R_{ij} + g_{ij} \nabla^2 - \nabla_i \nabla_j )(N \varphi \nabla^2 \varphi)\nb\\
                   &&      - \nabla_k \nabla_{(i} (N \varphi \nabla^k \nabla_{j)} \varphi) + \frac{1}{2}\nabla^2 (N \varphi \nabla_i \nabla_j \varphi) \nb\\
                   &&      + \frac{g_{ij}}{2} \nabla^\alpha \nabla^\beta ( N \varphi \nabla_\alpha \nabla_\beta \varphi)\nb\\
                   &&      - {\cal{G}}_{k (i} \nabla^k (N \varphi \nabla_{j)}\varphi ) + \frac{1}{2} {\cal{G}}_{ij} \nabla^k (N \varphi \nabla_k \varphi)\bigg\},\nb\\
(F_3^\varphi)_{ij} &=&    - \frac{1}{2}g_{ij} \varphi {\cal{G}}^{mn} a_m\nabla_n \varphi \nb\\
                   &&    -\varphi ( a_{(i} R_{j)k} \nabla^k \varphi + a^k R_{k(i} \nabla_{j)} \varphi)\nb\\
                   &&      + \frac{1}{2} (R - 2 \Lambda_g)  \varphi a_{(i} \nabla_{j)} \varphi \nb\\
                   &&      - \frac{1}{N}\bigg\{ - \frac{1}{2} (R_{ij} + g_{ij} \nabla^2 - \nabla_i \nabla_j ) (N \varphi a^k \nabla_k \varphi)\nb\\
                   &&     - \frac{1}{2} \nabla^k \Big[  \nabla_{(i} (\nabla_{j)} \varphi N \varphi)+ \nabla_{(i} (a_{j)} \varphi N \nabla_k \varphi) \Big] \nb\\
                   &&     + \frac{1}{2}\nabla^2 (N \varphi a_{(i} \nabla_{j)} \varphi) \nb\\
                   &&+ \frac{g_{ij}}{2} \nabla^\alpha \nabla^\beta (N \varphi a_\alpha \nabla_\beta \varphi)\bigg\}, \nb\\ 
(F_4^\varphi)_{ij} &=&    - \frac{1}{2}g_{ij} \hat{{\cal{G}}}^{mnkl}K_{mn} a_{(k}\nabla_{l)}\varphi \nb\\
                   &&    + \frac{1}{2 \sqrt{g} N} \partial_t [\sqrt{g} {\cal{G}}_{ij}^{\;\;k l} a_{(l} \nabla_{k)} \varphi] \nb\\
                   &&      + \frac{1}{2N} \nabla^\alpha \Big[  a_\alpha N_{(i} \nabla_{j)} \varphi +  N_{(i} a_{j)} \nabla_\alpha  \varphi \nb\\
                   &&      -  N_\alpha a_{(i} \nabla_{j)} \varphi + 2 g_{ij} N_\alpha a^k \nabla_k \varphi \Big] \nb\\
                   &&      + \frac{1}{N} \nabla_{(i} (N N_{j)} a^k \nabla_k \varphi)\nb\\
                   &&      + a^k K_{k(i} \nabla_{j)} \varphi + a_{(i} K_{j)k} \nabla^k \varphi \nb\\
                   &&      - K a_{(i} \nabla_{j)} \varphi - K_{ij} a^k \nabla_k \varphi, \nb\\
(F_5^\varphi)_{ij} &=&     -\frac{1}{2} g_{ij} \hat{{\cal{G}}}^{mnkl}[a_{(k}\nabla_{l)}\varphi][\nabla_m\nabla_n\varphi]\nb\\
                    && -a_{(i} \nabla^k \nabla_{j)} \varphi \nabla_k \varphi - a_k \nabla^k \nabla_{(i} \varphi \nabla_{j)}\varphi\nb\\
                   &&     + a_{(i} \nabla_{j)} \varphi \nabla^2 \varphi + a^k \nabla_k \varphi \nabla_i\nabla_j \varphi \nb\\
                   &&     + \frac{1}{2N} \bigg\{ \nabla^k (N \varphi a_k \nabla_i \varphi \nabla_j \varphi) \nb\\
                   &&   - 2 \nabla_{(i} (N \nabla_{j)} \varphi a^k
                          \nabla_k \varphi) \nb\\
                   &&     +g_{ij} \nabla^\alpha (\nabla_\alpha \varphi a^k \nabla_k \varphi)\bigg\}, \nb\\
(F_6^\varphi)_{ij} &=&   - \frac{1}{2} g_{ij}\hat{{\cal{G}}}^{mnkl} [a_{(m}\nabla_{n)}\varphi][a_{(k}\nabla_{l)}\varphi]\nb\\
                     &&-\frac{1}{2} (a^k \nabla_i \varphi- a_i \nabla^k \varphi) (a_k \nabla_j \varphi - a_j \nabla_k \varphi), \nb\\
(F_7^\varphi)_{ij} &=&   -\frac{1}{2} g_{ij}   \hat{{\cal{G}}}^{mnkl} [\nabla_{(n}\varphi][a_{m)(k}][\nabla_{l)}\varphi]\nb\\
                    && -\frac{1}{2} a_k^k \nabla_i \varphi \nabla_j \varphi - \frac{1}{2} a_{ij} \nabla^k \varphi \nabla_k \varphi \nb\\
                     &&     +  a^k_{(i} \nabla_{j)} \varphi \nabla_k \varphi - \frac{1}{2N}\bigg \{- \nabla_{(i} (N a_{j)} \nabla_k \varphi \nabla^k \varphi) \nb\\ &&+ \nabla^k (N a_{(i} \nabla_{j)} \varphi \nabla_k \varphi)\nb\\
                   &&      + \frac{g_{ij}}{2} \nabla^k (N a_k \nabla^m \varphi \nabla_m \varphi) \nb\\
                   &&- \frac{1}{2}\nabla^k (N a_k \nabla_i \varphi \nabla_j \varphi)\bigg\}, \nb\\
(F_8^\varphi)_{ij} &=&    - \frac{1}{2} g_{ij} (\nabla^2 \varphi+a_k\nabla^k\varphi)^2\nb\\
                   &&    -2 (\nabla^2 \varphi + a_k \nabla^k \varphi) (\nabla_i \nabla_j \varphi + a_i \nabla_j \varphi ) \nb\\
                   &&      -\frac{1}{N} \bigg \{ - 2 \nabla_{(j} [N \nabla_{i)} \varphi (\nabla^2 \varphi + a_k \nabla^k \varphi)] \nb\\
                   &&       + g_{ij} \nabla^\alpha [N (\nabla^2 \varphi + a_k \nabla^k \varphi) \nabla_\alpha \varphi]\bigg\}, \nb\\
(F_9^\varphi)_{ij} &=&    - \frac{1}{2} g_{ij}(\nabla^2 \varphi+a_k\nabla^k\varphi)K \nb\\
                    && - (\nabla^2 \varphi + a_k \nabla^k \varphi) K_{ij} \nb\\
                   &&     - (\nabla_i \nabla_j \varphi + a_i \nabla_j \varphi ) K \nb\\
                  &&     +\frac{1}{2 \sqrt{g} N}  \partial_t [\sqrt{g} (\nabla^2 \varphi + a_k \nabla^k \varphi) g_{ij}] \nb\\
                    &&      - \frac{1}{N}\bigg\{ - \nabla_{(j} [ N_{i)} (\nabla^2 \varphi + a_k \nabla^k \varphi)] \nb\\
                    &&     + \frac{1}{2} g_{ij} \nabla_\alpha [ N_\alpha (\nabla^2 \varphi + a_k \nabla^k \varphi) ]\nb\\
                    &&    -  \nabla_{(j} (N K \nabla_{i)} \varphi) + \frac{1}{2} g_{ij} \nabla_k (N K \nabla^k \varphi)
                            \bigg\}.\nb\\
\eqn



\begin{thebibliography}{nbound}

\bibitem{Horava}
P. Ho\v{r}ava, JHEP, {\bf 0903}, 020 (2009) [arXiv:0812.4287]; 
Phys. Rev. D{\bf 79}, 084008 (2009) [arXiv:0901.3775]; 
Phys. Rev.  Lett. {\bf 102}, 161301  (2009) [arXiv:0902.3657].

\bibitem{Lifshitz} E.M. Lifshitz, Zh. Eksp. Toer. Fiz., {\bf 11}, 255 (1941).

\bibitem{Visser} M. Visser, Phys. Rev. D{\bf 80}, 025011 (2009) 
[arXiv:0902.0590]; [arXiv:0912.4757]; 
C. Germani, A. Kehagias and K. Sfetsos, [arXiv:0906.1201].
  
\bibitem{BPS} D. Blas,  O. Pujolas and S. Sibiryakov, Phys. Rev. Lett. 
{\bf 104}, 181302 (2010) [arXiv:0909.3525]; 
JHEP, {\bf 1104}, 018 (2011) [arXiv.1007.3503].

\bibitem{KP}I. Kimpton and A. Padilla, J. High Energy Phys. 
{\bf 07}, 014 (2010) [arXiv:1003.5666].

\bibitem{LMP} H. L\"u, J. Mei and C.N. Pope, Phys. Rev. Lett. {\bf 103}, 
091301 (2009) [arXiv:0904.1595].

\bibitem{CalA} G. Calcagni, J. High Energy Phys., {\bf 09}, 112 (2009) 
[arXiv:0904.0829].

\bibitem{KK} R. G. Cai, L. M. Cao and N. Ohta, Phys. Rev. D{\bf 80}, 024003 
(2009) [arXiv:0904.3670]; 
A. Kehagias and K. Sfetsos, Phys. Lett. B{\bf 678}, 123 (2009) 
[arXiv:0905.0477]; 
M.-i. Park,  J. High Energy Phys. {\bf 09}, 123 (2009) [arXiv:0905.4480]; 
A. Ghodsi and E. Hatefi,  Phys. Rev. D{\bf 81}, 044016 (2010) [arXiv:0906.1237];
K. Izumi and S. Mukohyama, Phys. Rev. D{\bf 81}, 044008 (2010) 
[arXiv:0911.1814]; 
E.  Kiritsis, Phys. Rev. D{\bf 81}, 044009 (2010) [arXiv:0911.3164]; 
G. Koutsoumbas, E. Papantonopoulos, P. Pasipoularides and M. Tsoukalas, 
Phys. Rev. D{\bf 81}, 124014 (2010) [arxiv:1004.2289].

\bibitem{Hreview} P. Ho\v{r}ava, Class. Quantum Grav. {\bf 28}, 114012 (2011) 
[arXiv:1101.1081].

\bibitem{BLW} A. Borzou, K. Lin and A. Wang, JCAP, {\bf 05}, 006 (2011) 
[arXiv:1103.4366].

\bibitem{SVW} T. Sotiriou, M. Visser and S. Weinfurtner, Phys. Rev. Lett. 
{\bf 102},  251601 (2009) [arXiv:0904.4464]; 
J. High Energy Phys., {\bf 10}, 033 (2009) [arXiv:0905.2798].

\bibitem {KKa} E. Kiritsis and G. Kofinas, Nucl. Phys. B{\bf 821}, 467 (2009) 
[arXiv:0904.1334].

\bibitem{WM} A. Wang and R. Maartens, Phys. Rev. D {\bf 81}, 024009 (2010) 
[arXiv:0907.1748].

\bibitem{reviews}  A. Padilla,   J. Phys. Conf. Ser. {\bf 259}, 012033 (2010)  
[arXiv:1009.4074]; 
T.P. Sotiriou,  J. Phys. Conf. Ser. {\bf 283}, 012034 (2011) [arXiv:1010.3218]; 
T. Clifton, P.G. Ferreira, A. Padilla  and C. Skordis (2011) [arXiv:1106.2476].

\bibitem{Mukc} S. Mukohyama,  Class. Quantum Grav. {\bf 27}, 223101 (2010) 
[arXiv:1007.5199].

\bibitem{BS1} C. Bogdanos and E. N. Saridakis, Class. Quant. Grav. {\bf 27}, 
75005 (2010) [arXiv:0907.1636].

\bibitem{HWW} Y.-Q. Huang, A. Wang and Q. Wu, Mod. Phys. Lett. {\bf 25}, 2267
(2010) [arXiv:1003.2003].
 
\bibitem{WWa} A. Wang and Q. Wu, Phys. Rev. D{\bf 83}, 044025 (2011) 
[arXiv:1009.0268].

\bibitem{SC} C. Charmousis, G. Niz, A. Padilla and P.M. Saffin, JHEP, 
{\bf 08}, 070 (2009) [arXiv:0905.2579]; 
D. Blas, O. Pujolas and S. Sibiryakov, JHEP {\bf 10}, 029 (2009) 
[arXiv:0906.3046]; 
K. Koyama and F. Arroja, JHEP {\bf 03}, 061 (2010) [arXiv:0910.1998]; 
A. Papazoglou and T.P.  Sotiriou, Phys. Lett. B{\bf 685}, 197 (2010) 
[arXiv:0911.1299].

\bibitem{Vain} A.I. Vainshtein, Phys. Lett.  B {\bf 39}, 393 (1972); 
V.A.  Rubakov and P.G. Tinyakov, Phys. -Uspekhi, {\bf 51}, 759 (2008); K. 
Hinterbichler (2011) [arXiv:1105.3735].

\bibitem{Izumi:2011eh} K.~Izumi and S.~Mukohyama (2011) [arXiv:1105.0246].

\bibitem{GMW} A.E. Gumrukcuoglu, S. Mukohyama and A. Wang, [arXiv:1109.2609].

\bibitem{ZWWS} T. Zhu, Q. Wu, A. Wang and F.-W. Shu, Phys. Rev. D{\bf 84}, 
101502 (R) (2011) [arXiv:1108.1237].

\bibitem{ZSWW} T. Zhu, F.-W. Shu, Q. Wu and A. Wang, Phys. Rev. D{\bf 85}, 
044053 (2012) [arXiv: 1110.5106].

\bibitem{HMT} P. Ho\v{r}ava and C.M. Melby-Thompson, Phys. Rev. D{\bf 82}, 
064027 (2010) [arXiv:1007.2410].

\bibitem{WW} A. Wang and Y. Wu, Phys. Rev. D{\bf 83}, 044031 (2011) 
[arXiv:1009.2089].

\bibitem{HW} Y.-Q. Huang and A. Wang, Phys. Rev. D{\bf 83}, 104012 (2011) 
[arXiv:1011.0739].

\bibitem{Silva} A.M. da Silva, Class. Quantum Grav. {\bf 28}, 055011 (2011) 
[arXiv:1009.4885].

\bibitem{Kluson2} J.~Kluson, Phys. Rev. D{\bf 83}, 044049 (2011) 
[arXiv:1011.1857].

\bibitem{LWWZ}  K. Lin, A. Wang, Q. Wu and T. Zhu, Phys. Rev. D{\bf 84}, 
044051 (2011) [arXiv:1106.1486]. 

\bibitem{Lin2013} K. Lin and A. Wang, Phys. Rev. D{\bf 87}, 084041 (2013).

\bibitem{BS09} I. Bengtsson and J.M.M. Senovilla, Phys. rev. D{\bf 79}, 024027 
(2009).

\bibitem{HWWb} Y.-Q. Huang, A. Wang, and Q. Wu (2012) [arXiv:1201.4630].

\bibitem{BSW}  J. J. Greenwald, V.H. Satheeshkumar and A. Wang, JCAP, 
{\bf 12} , 007 (2010) [arXiv:1010.3794];
J. Greenwald, J. Lenells, J. X. Lu, V. H. Satheeshkumar, and A.  Wang,  
Phys. Rev. D{\bf 84}, 084040 (2011) [arXiv:1105.4259];
A.  Borzou, K. Lin, and A. Wang,   JCAP, {\bf 02}, 025 (2012) [arXiv:1110.1636].

\bibitem{AP} J.  Alexandre and P. Pasipoularides,   Phys. Rev. D
{\bf 83}, 084030 (2011) [arXiv:1010.3634];   
{\em ibid.}, D{\bf 84}, 084020 (2011) [arXiv:1108.1348].




\bibitem{Lin:2012bs}   K.~Lin, S.~Mukohyama and A.~Wang,
  Phys.\ Rev.\ D {\bf 86}, 104024 (2012).

\bibitem{Lin2014}   K.~Lin, S.~Mukohyama, A.~Wang and T. Zhu,
  Phys.\ Rev.\ D {\bf 89}, 084022 (2014).

\bibitem{ADM} C.W. Misner, K.S. Thorne and J.A. Wheeler, {\em Gravitation } 
(W.H. Freeman and Company, San Francisco, 1973), p. 484-528.

\bibitem{Goldoni2014} O. Goldoni, M.F.A. da Silva, G. Pinheiro and R. Chan,
Int. J. Mod. Phys. D, {\bf 23},  1450068 (2014).

\end{thebibliography}
\end{document}